\documentclass[review,1p]{elsarticle}
\usepackage{graphicx}
\usepackage{amssymb,amsthm,amsmath}
\usepackage{xspace}
\usepackage{epic,eepic}
\usepackage[usenames,dvipsnames]{pstricks}
\usepackage{epsfig}
\usepackage{pst-grad} 
\usepackage{pst-plot} 
\usepackage[normalem]{ulem}
\newtheorem{thm}{Theorem}[section]

\newtheorem{lem}[thm]{Lemma}

\theoremstyle{definition}

\theoremstyle{remark}
\newtheorem{rem}[thm]{Remark}

\newenvironment{prf}{{\bf \noindent Proof } }{\hfill$\square$\\}

\numberwithin{equation}{section}

\newcommand{\ignore}[1]{}

\newcommand{\dm}{$\delta-$minimum edge-colouring \xspace}

\newcommand{\sct}{$\{\alpha,\beta,\gamma\}$}
\begin{document}
\begin{frontmatter}
\title{A new bound for parsimonious edge-colouring of graphs with maximum degree three}
\author[lifo]{J-L. Fouquet}
\author[lifo]{J-M. Vanherpe}
\address[lifo]{L.I.F.O., Facult\'e des Sciences, B.P. 6759 \\
Universit\'e d'Orl\'eans, 45067 Orl\'eans Cedex 2, FR}

\date{\today}
\begin{abstract}
In a graph $G$ of maximum degree $3$, let $\gamma(G)$ denote the
largest fraction of edges that can be $3$ edge-coloured. Rizzi
\cite{Riz09} showed that $\gamma(G) \geq 1-\frac{2 \strut}{\strut 3
g_{odd}(G)}$ where $g_{odd}(G)$ is the odd girth of $G$, when $G$ is
triangle-free. In \cite{FouVan10a} we extended that  result to graphs
with maximum degree $3$. We show here that $\gamma(G) \geq 1-\frac{2
\strut}{\strut 3 g_{odd}(G)+2}$, which leads to $\gamma(G) \geq
\frac{15}{17}$ when considering graphs with odd girth at least $5$,
distinct from the Petersen graph.
\ignore{As a corollary we get also this lower bound $\frac{15}{17}$ for bridgeless cubic graphs distinct from the Petersen graph.}

\end{abstract}
\end{frontmatter}
\section{Introduction}
Throughout this paper, we shall be concerned with connected graphs
with maximum degree $3$. Staton \cite{Sta80} (and independently Locke \cite{Loc82}) showed that
whenever $G$ is a cubic graph distinct from $K_4$ then $G$ contains
a bipartite subgraph (and hence a $3$-edge colourable graph, by
K\"{o}nig's theorem \cite{Kon16}) with at least $\frac{7}{9}$ of the
edges of $G$. Bondy and Locke \cite{BonLock86} obtained $\frac{4}{5}$
when considering graphs with maximum degree at most $3$.

In \cite{AlbHaa96} Albertson and Haas showed that whenever $G$ is a
cubic graph, we have $\gamma(G) \geq \frac{13}{15}$ (where
$\gamma(G)$ denote the largest fraction of edges of $G$ that can be
$3$ edge-coloured) while for graphs with maximum degree $3$ they
obtained $\gamma(G) \geq \frac{26}{31}$. Steffen \cite{Ste04} proved
that the only cubic bridgeless graph with $\gamma(G) =
\frac{13}{15}$ is the Petersen graph. In \cite{FouVan10a}, we
extended  this result to graphs with maximum degree $3$ where
bridges are allowed. With the exception of $G_5$ (a $C_{5}$ with two
chords), the graph $P^{'}$ obtained from two copies of $G_5$ by
joining by an edge the two vertices of degree $2$ and the Petersen
graph, every graph $G$ is such that $\gamma(G) > \frac{13}{15}$.
Rizzi \cite{Riz09} showed that $\gamma(G) \geq 1-\frac{2 \strut
}{\strut 3 g_{odd}(G)}$ where $g_{odd}(G)$ is the odd girth of $G$,
when $G$ is triangle-free. In \cite{FouVan10a} we extended that
result to graph with maximum degree $3$ (triangles are allowed). We
show here that $\gamma(G) \geq 1-\frac{2 \strut}{\strut3
g_{odd}(G)+2}$, which leads to $\gamma \geq \frac{15}{17}$ when
considering graphs with odd girth at least $5$, distinct from the
Petersen graph.

\begin{thm} \label{Theorem:Principal}let $G$ be a graph with maximum degree $3$ distinct from the Petersen graph.
Then $\gamma(G) \geq 1-\frac{2 \strut}{\strut 3 g_{odd}(G)+2}$
\end{thm}

\section{Technical lemmas}
Let $\phi :\ E(G) \rightarrow \{\alpha,\beta,\gamma,\delta\}$ be a
proper edge-colouring of $G$. It is often of interest to try to use
one colour (say $\delta$) as few as possible. When an edge colouring
is optimal, following this constraint, we shall say that $\phi$ is
$\delta-minimum$.  Since any two \dm of $G$ have the same number of
edges coloured $\delta$ we shall denote by $s(G)$ this number. For
$x,y \in \{\alpha,\beta,\gamma,\delta\}$, $x\neq y$, $\phi(x,y)$ is
the partial subgraph of $G$ spanned by these two colours (this
subgraph being a union of paths and even cycles where the colours
$x$ and $y$ alternate) and $E_{\phi}(x)$ is the set of edges
coloured with $x$.

In \cite{Fou80} we gave without proof (in French, see \cite{FouVan10b} for
a translation) results on $\delta$-minimum edge-colourings of cubic graphs

\begin{lem} \cite{Fou80,FouPhd,FouVan10b}\label{Lemma:OddCycleAssociated} Let $\phi$ be a \dm of $G$.
For any edge $e=uv$ coloured with $\delta$ there are two colours $x$
and $y$ in \sct such that the connected component of $\phi(x,y)$
containing the two ends of $e$ is an even  path joining these two
ends. Moreover $e$ has one end of degree $2$ and the other of degree $3$ or the two
ends of degree $3$
\end{lem}

\begin{rem} \label{Remark:NoUnicityFondamentalOddCycle} An edge coloured with $\delta$ by the \dm $\phi$  is in $A_{\phi}$ when its ends
can be connected by a path of $\phi(\alpha,\beta)$, $B_{\phi}$ by a
path of $\phi(\beta,\gamma)$ and $C_{\phi}$   by a path of
$\phi(\alpha,\gamma)$.
It is clear that $A_{\phi}$, $B_{\phi}$ and
$C_{\phi}$ are not necessarily pairwise disjoint since an edge
coloured with $\delta$ with one end of degree $2$ is contained in $2$
such sets. Assume indeed that $e=uv$ is coloured with $\delta$ while
$d(u)=3$ and $d(v)=2$ then, if $u$ is incident to $\alpha$ and
$\beta$ and $v$ is incident to $\gamma$ we have an alternating path
whose ends are $u$ and $v$ in $\phi(\alpha,\gamma)$ as well as in
$\phi(\beta,\gamma)$. Hence $e$ is in $A_{\phi} \cap B_{\phi}$.
When $e \in A_{\phi}$  we can associate to $e$ the odd cycle
$C_{A_{\phi}}(e)$  obtained by considering the
path of $\phi(\alpha,\beta)$  together with $e$. We define in the
same way $C_{B_{\phi}}(e)$ and $C_{C_{\phi}}(e)$ when $e$ is in
$B_{\phi}$ or $C_{\phi}$. In the following lemma we consider an edge
in $A_{\phi}$, an analogous result  holds true whenever we consider
edges in $B_{\phi}$ or $C_{\phi}$ as well.
\end{rem}

\begin{lem} \cite{Fou80,FouPhd,FouVan10b}\label{Lemma:FondamentalOddCycle}
Let $\phi$ be a \dm of $G$ and let $e$ be an edge in $A_{\phi}$ then for any edge $e' \in C_{A_{\phi}}(e)$
there is a \dm $\phi'$ such that
$E_{\phi'}(\delta)=E_{\phi}(\delta)-\{e\}\cup \{e'\}$, $e' \in A_{\phi'}$ and
$C_{A_{\phi}}(e)=C_{A_{\phi'}}(e')$. Moreover, each edge outside
$C_{A_{\phi}}(e)$ but incident with this cycle is coloured $\gamma$, $\phi$ and $\phi'$ only differ on the edges of $C_{A_{\phi}}(e)$.
\end{lem}

For each edge $e \in E_{\phi}(\delta)$ (where $\phi$ is a \dm of
$G$) we can associate one or two odd cycles following the fact
that $e$ is in one or two sets among $A_{\phi}$, $B_{\phi}$ or
$C_{\phi}$. Let ${\mathcal C}$ be the set of odd cycles associated
to edges in $E_{\phi}(\delta)$.

\begin{lem}\cite{Fou80,FouPhd,FouVan10b} \label{Lemma:DisjointOddCycles}
Let $e_1,e_2 \in E_{\phi}(\delta)$  and let $C_1,C_2 \in \mathcal C$
be such that $C_1\neq C_2$, $e_1 \in E(C_1)$ and $e_2 \in E(C_2)$
then $C_1$ and $C_2$ are (vertex) disjoint.
\end{lem}


\begin{lem}\cite{Fou80,FouPhd,FouVan10b}\label{Lemma:OneVertexInNeighborhood}
Let $e_1=uv_1$ be an edge of $E_{\phi}(\delta)$ such that $v_1$ has
degree $2$ in $G$. Then $v_1$ is the only vertex in $N(u)$ of degree
 $2$ and for any edge $e_2=u_2v_2 \in E_{\phi}(\delta)$,  $\{e_1,e_2\}$ induces a $2K_2$.
\end{lem}

\begin{lem}\cite{Fou80,FouPhd,FouVan10b}\label{Lemma:2K2+SinonAuPlusUneArete}
Let $e_1$ and $e_2$ be two edges of $E_{\phi}(\delta)$.
If $e_1$ and $e_2$ are contained in two distinct sets of $A_{\phi},B_{\phi}$ or $C_{\phi}$ then $\{e_1,e_2\}$ induces a $2K_2$ otherwise $e_1,e_2$ are joined by at most one edge.
\end{lem}

\begin{lem} \cite{Fou80,FouPhd,FouVan10b}\label{Lemma:Oneedgewiththree}
Let $e_1, e_2$ and $e_3$ be three distinct edges of
$E_{\phi}(\delta)$ contained in the same  set $A_{\phi},B_{\phi}$ or
$C_{\phi}$. Then $\{e_1,e_2,e_3\}$  induces a subgraph with at most
four edges.
\end{lem}

\begin{lem} \cite{FouVan10a} \label{Lemma:bG} Let $G$ be a graph with maximum degree $3$ then $\gamma(G)
= 1-\frac{s(G)}{m}$.
\end{lem}

A cubic graph $G$ on $n$ vertices is called a {\em permutation graph} if $G$ has a
perfect matching $M$ such $G-M$ is the union of two chordless cycles
$A$ and $B$ of equal length $\frac{n}{2}$. When we delete one edge of the perfect matching $M$ in the above permutation graph $G$, we shall say that the graph obtained is a {\em near-permutation graph}.
\begin{lem} \label{Lemma:NearPermutation}
Let $G$ be a permutation graph or a near-permutation graph with $n$ vertices and odd girth $\frac{n}{2}$. Suppose that $G$ is not $3$-edge colourable. Then $G$ is the Petersen graph or the Petersen graph minus one edge.
\end{lem}

\begin{prf} Obviously, since $G$ is not $3$-edge colourable,   $\frac{n}{2}$ is certainly odd.
Let $A=a_{0}a_{1} \ldots a_{2k}$ and $B=b_{0}b_{1} \ldots b_{2k}$ be
the two chordless cycles of length $\frac{n}{2}=2k+1$ which
partition $V(G)$. When $2k+1=3$, it can be easily verified that $G$
is $3-$edge colourable and when $2k+1=5$, $G$ is the Petersen graph
or the Petersen graph minus one edge. Assume thus that $2k+1\geq 7$.

Since at most one vertex of $A$ and one vertex of $B$ have degree
$2$, we can suppose that $a_{0}, a_{1}, a_{2}$ are joined to $3$
distinct vertices of $B$. Without loss of generality we suppose that
$a_{0}b_{0} \in E(G)$. Since $G$ is not $3-$edge colourable
$a_{1}b_{1}$ and $a_{1}b_{2k}$ are not edges of $G$ and since the
odd girth is at least $7$ we do not have the edges $a_{1}b_{2}$ and
$a_{1}b_{2k-1}$. Henceforth let $b_{i}$ ($2 < i\leq 2k-2$) the
neighbour of $a_{1}$. One of the two paths determined by $b_{0}$ and
$b_{i}$ on $B$ must have odd length. Suppose, without loss of
generality,  that $b_{0}b_{1} \ldots b_{i}$ has odd length then we
have an odd cycle $a_{0}b_{0}b_{1} \ldots b_{i}a_{1}$ whose length
is at least $2k+1$, which leads to $i=2k-2$.

The vertex $a_{2}$ is not joined to $b_{2k-1}$ or $b_{2k-3}$,
otherwise $G$ is $3$-edge colourable, neither to $b_{2k}$
,$b_{2k-4}$ or $b_{1}$, otherwise the odd girth is $5$. Henceforth
$a_{2}$ is joined to some vertex $b_{j}$ with $2 \leq j \leq 2k-5$.
If $j$ is odd then $b_{0}b_{1} \ldots b_{j} a_{2}a_{1}a_{0}$ is an
odd cycle of length at most $2k-1$, impossible. If $j$ is even then
$b_{j}\ldots b_{2k-3}b_{2k-2}a_{1}a_{2}$ an odd cycle of length at
most $2k-1$, impossible.
\end{prf}

\section{Proof of Theorem \ref{Theorem:Principal}}

\begin{prf}

Let $\phi$ be a \dm of $G$ and $E_{\phi}(\delta)=\{e_1,e_2\ldots e_{s(G)}\}$. ${\mathcal C}$ being the set of odd cycles associated to edges
in $E_{\phi}(\delta)$, we write $\mathcal C=\{C_1,C_2\ldots C_{s(G)}\}$ and suppose that for $i=1,2\ldots s(G)$, $e_i$ is an edge of $C_i$.
We know by Lemma \ref{Lemma:DisjointOddCycles} that the cycles of $\mathcal C$ are vertex-disjoint.

Let $\displaystyle l(\mathcal C) = \sum_{C \in \mathcal C} l(C)$
(where $l(C)$ is the length of the cycle $C$) and assume that $\phi$
has been chosen in such a way that $l(\mathcal C)$ is maximum.

Let us write $\mathcal C=\mathcal C_2 \cup \mathcal C_3$, where $\mathcal C_2$ denotes the set of odd
cycles of $\mathcal C$ which have a
vertex of degree $2$, while $\mathcal C_3$ is for the set of cycles in $\mathcal C$ whose all vertices
have degree $3$. Let $k=|\mathcal C_2|$,
obviously we have $0\leq k\leq s(G)$ and $\mathcal C_2\cap\mathcal C_3=\emptyset$.

If $C_i\in\mathcal C_2$, we may suppose that $e_i$ has a vertex of degree $2$ (see Lemma \ref{Lemma:FondamentalOddCycle}) and we
can associate to $e_i$ another odd cycle say $C^{'}_i$ (Remark \ref{Remark:NoUnicityFondamentalOddCycle}) whose edges distinct from
 $e_i$ form  an even path $P_{i}$ using at least $\frac{g_{odd}(G)}{2}$ edges which are not edges of $C_i$.
When $l(C_{i}) = g_{odd}(G)$, $C_{i}$ has no chord and it is an easy
task to find a supplementary edge of $P_{i}$ not belonging to
$C_{i}$. When $l(C_{i})\geq g_{odd}(G)+2$ (recall that $C_{i}$ has
odd length), we can choose an edge of  $C_{i}$ as a {\em private}
edge. In both cases
 $C_i\cup C^{'}_i$ contains at least $\frac{3}{2}g_{odd}(G)+1$ edges. Consequently there are at least $k\times (\frac{3}{2} g_{odd}(G) + 1)$
 edges in $\displaystyle\bigcup_{C_i\in\mathcal C_2}(C_i\cup C^{'}_i)$.

When $C_i\in\mathcal C_3$, $C_i$ contains at least $g_{odd}(G)$ edges, moreover, each vertex of $C_i$ being of degree $3$, there are
$\frac{s(G)-k}{2}\times g_{odd}(G)$ additional edges which are incident to a vertex of $\displaystyle\bigcup_{C_i\in\mathcal C_3} C_{i}$.
Let us remark that each above additional edge is counted as $\frac{1}{2}$ whatever are these edges. In order to refine our counting,
we need to introduce the following notion of {\em free} edge. An edge will be said to be {\em free} when at most one end belongs to
some $C_i\in\mathcal C_3$.

Suppose that we can associate to each $C_i \in\mathcal C_3$ one
private edge or two private free edges. Since $C_i\cap
C_j=\emptyset$ and $C^{'}_i\cap C_j=\emptyset$ ($1\leq i,j\leq
s(G)$, $i\neq j$), we would have
$$m\geq (k \times (\frac{3}{2}g_{odd}(G)+1) +(s(G)-k)\times (g_{odd}(G)+1)+\frac{s(G)-k}{2}\times (g_{odd}(G))$$
and
$$m \geq  s(G)\times (\frac{3}{2}\times g_{odd}(G)+1).$$

Consequently $\gamma(G)= 1-\frac{s(G)}{m}\geq 1-\frac{2
\strut}{\strut 3g_{odd}(G)+2}$, as claimed.

Our goal now is to associate to each $C_i \in\mathcal C_3$ one
private edge or two private free edges.

When $C_{i}$ is incident to at least two free edges, let us choose any two such free edges as the private free edges associate
to $C_{i}$. By definition, these two free edges are not incident to any $C_j \in\mathcal C_3$ with $j \neq i$, insuring thus
that they cannot be associated to $C_{j}$ . When $l(C_{i}) \geq g_{odd}(G)+2$, we choose any edge of $C_{i}$ as a private edge.

Assume thus that $l(C_{i})=g_{odd}(G)$. Hence $C_{i}=x_{0}x_{1} \ldots x_{g_{odd}(G)-1}$ is chordless. Suppose that $C_{i}$
is incident to at most one free edge. Without loss of generality, we can consider that $x_{0}$ is the only possible vertex
of $C_{i}$ incident to some free edge. Since the edge incident to $x_{1}$, not belonging to $C_{i}$, is not free,
let $C_{j} \in \mathcal C_{3}$, $i \neq j$, such that $x_{1}$ is adjacent to $y_{1} \in C_{j}$. In the same way, the edge
incident to $x_{2}$, not belonging to $C_{i}$, is not free. Let $C_{j^{'}}\in \mathcal C_{3}$, $i \neq j^{'}$, such that $x_{2}$
is adjacent to $z_{2} \in C_{j^{'}}$.  Suppose that $j \neq j^{'}$, then by Lemma \ref{Lemma:FondamentalOddCycle} we can consider
that $x_{1}x_{2}$ is coloured $\delta$ by $\phi$ as well as one of the edges of $C_{j}$ incident with the vertex $y_{1}$  and
one of the edges of $C_{j^{'}}$ incident with $z_{2}$, a contradiction with Lemma \ref{Lemma:Oneedgewiththree} or
Lemma \ref{Lemma:2K2+SinonAuPlusUneArete}.
Henceforth, $x_{2}$ is adjacent to some vertex of $C_{j}$. In the same way every vertex of $C_{i}$, distinct from $x_{0}$,
is adjacent to some vertex of $C_{j}$.

In this situation, let us say that $C_{i}$ is {\em extremal for
$C_{j}$}. Assume that we have a set of $p \geq 2$ distinct extremal
cycles of $\mathcal C_{3}$ for $C_{j}$. Since two distinct extremal
cycles have no consecutive neighbours in $C_{j}$ (otherwise we get a
contradiction with Lemmas \ref{Lemma:Oneedgewiththree} or
\ref{Lemma:2K2+SinonAuPlusUneArete} as above), $C_{j}$ has length at
least $p \times (g_{odd}(G)-1)+2$. But $p \times (g_{odd}(G)-1)+2
\geq g_{odd}(G)+p+1$ as soon as $g_{odd}(G) \geq 3$. hence, when $p
\geq 2$, we can associate to each extremal cycle for $C_{j}$ a
private edge belonging to $C_{j}$ as well as a private edge for
$C_{j}$ itself, since $l(C_{j}) \geq g_{odd}(G)+p+1$.

Assume thus that $C_{i}$ is the only extremal cycle for $C_{j}$. If $l(C_{j}) \geq g_{odd}(G)+2$, we can associate any edge
of $C_{j}$ as a private edge of $C_{i}$ and any other edge of $C_{j}$ as a private edge of $C_{j}$. It remains thus to consider
the case where $l(C_{j})=g_{odd}(G)$. In that situation, $C_{i}$ is  an extremal cycle for $C_{j}$, as well as $C_{j}$ is  an
extremal cycle for $C_{i}$ and the subgraph induced by $C_{i} \cup C_{j}$ is a permutation graph or a near permutation graph
with odd girth $g_{odd}(G)=  \frac{\mid C_{i} \cup C_{j} \mid}{2}$.

By Lemma \ref{Lemma:NearPermutation}, $G$ itself is the Petersen graph or $C_{i} \cup C_{j}$ induces a Petersen graph minus
one edge (by the way, $g_{odd}(G)=5$). By hypothesis, the first case is excluded. Assume thus that $C_{i} \cup C_{j}$ induces
a Petersen graph minus one edge. In the last part of this proof, we show that this situation is not possible.

In order to fix the situation let $H$ be the subgraph of $G$ not
containing $C_{i} \cup C_{j}$. We suppose that $C_{j}$ is the
chordless cycle of length $5$ $y_{0}y_{3}y_{1}y_{4}y_{2}$ while
$x_{1}y_{1},x_{2}y_{2},x_{3}y_{3}$ and $x_{4}y_{4}$ are the edges
joining $C_{i}$ to $C_{j}$. Moreover $x_{0}$ is joined to some
vertex $a \in V(H)$ and $y_{0}$ is joined to some vertex $b \in
V(H)$. Without loss of generality, we can consider that $\phi$
colours alternately the edges of $C_{i}$ ($C_{j}$ respectively) with
$\beta$ and $\gamma$ with the exception of the edge $x_{0}x_{1}$
coloured with $\delta$ ($y_{0}y_{3}$ respectively).

The edges $x_{1}y_{1},x_{2}y_{2},x_{3}y_{3}$ and $x_{4}y_{4}$ are
thus coloured with $\alpha$ as well as the edges $x_{0}a$ and
$y_{0}b$ (let us remark that $a \neq b$). The final situation is
depicted in Figure \ref{Figure:FinalSituation_1}.

\begin{figure}[htb]
\begin{center}

\begin{center}
\scalebox{1} 
{
\begin{pspicture}(0,-2.598125)(12.012813,2.598125)
\definecolor{color414b}{rgb}{0.03137254901960784,0.023529411764705882,0.023529411764705882}
\pscircle[linewidth=0.04,dimen=outer,fillstyle=solid,fillcolor=color414b](1.44,1.1996875){0.1}
\pscircle[linewidth=0.04,dimen=outer,fillstyle=solid,fillcolor=color414b](1.44,-0.8203125){0.1}
\pscircle[linewidth=0.04,dimen=outer,fillstyle=solid,fillcolor=color414b](3.44,1.1996875){0.1}
\pscircle[linewidth=0.04,dimen=outer,fillstyle=solid,fillcolor=color414b](3.46,-0.7803125){0.1}
\pscircle[linewidth=0.04,dimen=outer,fillstyle=solid,fillcolor=color414b](5.42,1.2196875){0.1}
\pscircle[linewidth=0.04,dimen=outer,fillstyle=solid,fillcolor=color414b](7.44,1.1996875){0.1}
\pscircle[linewidth=0.04,dimen=outer,fillstyle=solid,fillcolor=color414b](9.44,1.1996875){0.1}
\pscircle[linewidth=0.04,dimen=outer,fillstyle=solid,fillcolor=color414b](11.46,1.2396874){0.1}
\pscircle[linewidth=0.04,dimen=outer,fillstyle=solid,fillcolor=color414b](5.44,-0.8003125){0.1}
\pscircle[linewidth=0.04,dimen=outer,fillstyle=solid,fillcolor=color414b](7.44,-0.8003125){0.1}
\pscircle[linewidth=0.04,dimen=outer,fillstyle=solid,fillcolor=color414b](9.42,-0.7803125){0.1}
\pscircle[linewidth=0.04,dimen=outer,fillstyle=solid,fillcolor=color414b](11.42,-0.7603125){0.1}
\psellipse[linewidth=0.04,dimen=outer](1.15,0.1096875)(1.15,2.27)
\psline[linewidth=0.03cm,fillcolor=color414b](1.4,1.1996875)(11.46,1.1996875)
\psline[linewidth=0.03cm,fillcolor=color414b](1.4,-0.8003125)(11.46,-0.8003125)
\psline[linewidth=0.03cm,fillcolor=color414b](5.4,1.2196875)(7.44,-0.8203125)
\psline[linewidth=0.03cm,fillcolor=color414b](7.42,1.1996875)(11.4,-0.7803125)
\psline[linewidth=0.03cm,fillcolor=color414b](11.46,1.2596875)(9.4,-0.8403125)
\psline[linewidth=0.03cm,fillcolor=color414b](9.42,1.2196875)(5.46,-0.8003125)
\usefont{T1}{ptm}{m}{n}
\rput(3.3975,0.8396875){\small $x_0$}
\usefont{T1}{ptm}{m}{n}
\rput(5.4175,0.8596875){\small $x_1$}
\usefont{T1}{ptm}{m}{n}
\rput(7.4575,0.8796875){\small $x_2$}
\usefont{T1}{ptm}{m}{n}
\rput(9.4575,0.8596875){\small $x_3$}
\usefont{T1}{ptm}{m}{n}
\rput(11.5375,0.8596875){\small $x_4$}
\usefont{T1}{ptm}{m}{n}
\rput(3.4775,-0.3603125){\small $y_0$}
\usefont{T1}{ptm}{m}{n}
\rput(5.4375,-0.3803125){\small $y_3$}
\usefont{T1}{ptm}{m}{n}
\rput(7.6175,-0.4003125){\small $y_1$}
\usefont{T1}{ptm}{m}{n}
\rput(9.3175,-0.4003125){\small $y_4$}
\usefont{T1}{ptm}{m}{n}
\rput(11.5375,-0.3603125){\small $y_2$}
\psbezier[linewidth=0.03,fillcolor=color414b](3.52,1.2596875)(3.8,2.1596875)(6.22,2.2196875)(7.24,2.2196875)(8.26,2.2196875)(11.16,2.0796876)(11.46,1.2996875)
\psbezier[linewidth=0.03,fillcolor=color414b](3.4,-0.7203125)(3.72,-1.6803125)(6.200213,-1.8996695)(7.2,-1.9203125)(8.199787,-1.9409555)(11.16,-1.8803124)(11.42,-0.8203125)
\usefont{T1}{ptm}{m}{n}
\rput(0.6514062,0.1646875){$H$}
\usefont{T1}{ptm}{m}{n}
\rput(1.2214062,1.5846875){$a$}
\usefont{T1}{ptm}{m}{n}
\rput(1.3114063,-1.2753125){$b$}
\usefont{T1}{ptm}{m}{n}
\rput(2.7914062,1.5046875){$\alpha$}
\usefont{T1}{ptm}{m}{n}
\rput(2.7914062,-1.1553125){$\alpha$}
\usefont{T1}{ptm}{m}{n}
\rput(7.5114064,0.4646875){$\alpha$}
\usefont{T1}{ptm}{m}{n}
\rput(5.9714065,0.4446875){$\alpha$}
\usefont{T1}{ptm}{m}{n}
\rput(9.551406,0.4846875){$\alpha$}
\usefont{T1}{ptm}{m}{n}
\rput(6.411406,-1.0353125){$\beta$}
\usefont{T1}{ptm}{m}{n}
\rput(6.431406,1.4046875){$\beta$}
\usefont{T1}{ptm}{m}{n}
\rput(10.431406,1.3846875){$\beta$}
\usefont{T1}{ptm}{m}{n}
\rput(10.4114065,-1.0953125){$\beta$}
\usefont{T1}{ptm}{m}{n}
\rput(4.431406,1.3846875){$\delta$}
\usefont{T1}{ptm}{m}{n}
\rput(4.431406,-1.0753125){$\delta$}
\usefont{T1}{ptm}{m}{n}
\rput(7.6514063,-2.3753126){$\gamma$}
\usefont{T1}{ptm}{m}{n}
\rput(8.391406,-0.9953125){$\gamma$}
\usefont{T1}{ptm}{m}{n}
\rput(7.271406,2.4046874){$\gamma$}
\usefont{T1}{ptm}{m}{n}
\rput(8.4114065,1.3846875){$\gamma$}
\usefont{T1}{ptm}{m}{n}
\rput(11.191406,0.5046875){$\alpha$}
\end{pspicture} 
}
\end{center}
\end{center}
\caption{Final situation} \label{Figure:FinalSituation_1}
\end{figure}
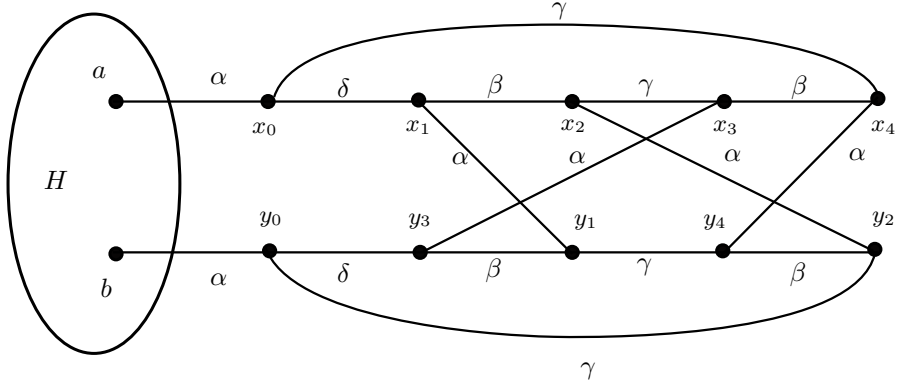

Without changing any colour in $H$ and keeping the colour $\alpha$ for the edges $x_{0}a$ and $y_{0}b$, we
can construct a new \dm $\phi^{'}$ in the following way (see Figure \ref{Figure:NewColouring}):
\begin{itemize}
  \item [$\bullet$] $x_{0}x_{4}$ and $x_{1}x_{2}$ are coloured with $\delta$
  \item [$\bullet$] $x_{4}y_{4},y_{0}y_{2},x_{2}x_{3}$ and $y_{1}y_{3}$ are coloured with $\beta$
  \item [$\bullet$] $x_{1}y_{1},x_{3}y_{3}$ and $y_{4}y_{2}$ are coloured with $\alpha$
  \item [$\bullet$] $x_{0}x_{1},x_{3}x_{4},y_{0}y_{3},y_{1}y_{4}$ and $x_{2}y_{2}$ are coloured with $\gamma$.
\end{itemize}

\begin{figure}[htb]
\begin{center}

\begin{center}
\scalebox{1} 
{
\begin{pspicture}(0,-2.5329688)(12.012813,2.5329688)
\definecolor{color414b}{rgb}{0.03137254901960784,0.023529411764705882,0.023529411764705882}
\pscircle[linewidth=0.04,dimen=outer,fillstyle=solid,fillcolor=color414b](1.44,1.1345313){0.1}
\pscircle[linewidth=0.04,dimen=outer,fillstyle=solid,fillcolor=color414b](1.44,-0.8854687){0.1}
\pscircle[linewidth=0.04,dimen=outer,fillstyle=solid,fillcolor=color414b](3.44,1.1345313){0.1}
\pscircle[linewidth=0.04,dimen=outer,fillstyle=solid,fillcolor=color414b](3.46,-0.84546876){0.1}
\pscircle[linewidth=0.04,dimen=outer,fillstyle=solid,fillcolor=color414b](5.42,1.1545312){0.1}
\pscircle[linewidth=0.04,dimen=outer,fillstyle=solid,fillcolor=color414b](7.44,1.1345313){0.1}
\pscircle[linewidth=0.04,dimen=outer,fillstyle=solid,fillcolor=color414b](9.44,1.1345313){0.1}
\pscircle[linewidth=0.04,dimen=outer,fillstyle=solid,fillcolor=color414b](11.46,1.1745312){0.1}
\pscircle[linewidth=0.04,dimen=outer,fillstyle=solid,fillcolor=color414b](5.44,-0.86546874){0.1}
\pscircle[linewidth=0.04,dimen=outer,fillstyle=solid,fillcolor=color414b](7.44,-0.86546874){0.1}
\pscircle[linewidth=0.04,dimen=outer,fillstyle=solid,fillcolor=color414b](9.42,-0.84546876){0.1}
\pscircle[linewidth=0.04,dimen=outer,fillstyle=solid,fillcolor=color414b](11.42,-0.8254688){0.1}
\psellipse[linewidth=0.04,dimen=outer](1.15,0.04453125)(1.15,2.27)
\psline[linewidth=0.03cm,fillcolor=color414b](1.4,1.1345313)(11.46,1.1345313)
\psline[linewidth=0.03cm,fillcolor=color414b](1.4,-0.86546874)(11.46,-0.86546874)
\psline[linewidth=0.03cm,fillcolor=color414b](5.4,1.1545312)(7.44,-0.8854687)
\psline[linewidth=0.03cm,fillcolor=color414b](7.42,1.1345313)(11.4,-0.84546876)
\psline[linewidth=0.03cm,fillcolor=color414b](11.46,1.1945312)(9.4,-0.90546876)
\psline[linewidth=0.03cm,fillcolor=color414b](9.42,1.1545312)(5.46,-0.86546874)
\usefont{T1}{ptm}{m}{n}
\rput(3.3975,0.77453125){\small $x_0$}
\usefont{T1}{ptm}{m}{n}
\rput(5.4175,0.7945312){\small $x_1$}
\usefont{T1}{ptm}{m}{n}
\rput(7.4575,0.81453127){\small $x_2$}
\usefont{T1}{ptm}{m}{n}
\rput(9.4575,0.7945312){\small $x_3$}
\usefont{T1}{ptm}{m}{n}
\rput(11.5375,0.7945312){\small $x_4$}
\usefont{T1}{ptm}{m}{n}
\rput(3.4775,-0.42546874){\small $y_0$}
\usefont{T1}{ptm}{m}{n}
\rput(5.4375,-0.44546875){\small $y_3$}
\usefont{T1}{ptm}{m}{n}
\rput(7.6175,-0.46546876){\small $y_1$}
\usefont{T1}{ptm}{m}{n}
\rput(9.3175,-0.46546876){\small $y_4$}
\usefont{T1}{ptm}{m}{n}
\rput(11.5375,-0.42546874){\small $y_2$}
\psbezier[linewidth=0.03,fillcolor=color414b](3.52,1.1945312)(3.8,2.0945313)(6.22,2.1545312)(7.24,2.1545312)(8.26,2.1545312)(11.16,2.0145311)(11.46,1.2345313)
\psbezier[linewidth=0.03,fillcolor=color414b](3.4,-0.78546876)(3.72,-1.7454687)(6.200213,-1.9648257)(7.2,-1.9854687)(8.199787,-2.0061116)(11.16,-1.9454688)(11.42,-0.8854687)
\usefont{T1}{ptm}{m}{n}
\rput(0.41140625,0.09953125){$H$}
\usefont{T1}{ptm}{m}{n}
\rput(1.2214062,1.5195312){$a$}
\usefont{T1}{ptm}{m}{n}
\rput(1.3114063,-1.3404688){$b$}
\usefont{T1}{ptm}{m}{n}
\rput(2.7914062,1.4395312){$\alpha$}
\usefont{T1}{ptm}{m}{n}
\rput(2.7914062,-1.2204688){$\alpha$}
\usefont{T1}{ptm}{m}{n}
\rput(7.5114064,0.39953125){$\alpha$}
\usefont{T1}{ptm}{m}{n}
\rput(5.9714065,0.37953126){$\alpha$}
\usefont{T1}{ptm}{m}{n}
\rput(6.411406,-1.1804688){$\beta$}
\usefont{T1}{ptm}{m}{n}
\rput(6.4514065,1.3995312){$\delta$}
\usefont{T1}{ptm}{m}{n}
\rput(8.311406,1.4195312){$\beta$}
\usefont{T1}{ptm}{m}{n}
\rput(11.191406,0.37953126){$\beta$}
\usefont{T1}{ptm}{m}{n}
\rput(4.6314063,1.4395312){$\gamma$}
\usefont{T1}{ptm}{m}{n}
\rput(9.631406,0.41953126){$\gamma$}
\usefont{T1}{ptm}{m}{n}
\rput(4.5314064,-1.2004688){$\gamma$}
\usefont{T1}{ptm}{m}{n}
\rput(7.0914063,2.3395312){$\delta$}
\usefont{T1}{ptm}{m}{n}
\rput(10.391406,1.3795313){$\gamma$}
\usefont{T1}{ptm}{m}{n}
\rput(10.291407,-1.1004688){$\alpha$}
\psbezier[linewidth=0.05,linestyle=dashed,dash=0.16cm 0.16cm,fillcolor=color414b](1.3668365,1.0945313)(0.94,0.9945313)(0.86,0.55510134)(0.82,0.17453125)(0.78,-0.2060388)(1.06,-0.90546876)(1.44,-0.8615522)
\usefont{T1}{ptm}{m}{n}
\rput(8.351406,-1.1804688){$\gamma$}
\usefont{T1}{ptm}{m}{n}
\rput(7.1914062,-2.3604689){$\beta$}
\end{pspicture} 
}
\end{center}
\end{center}
\caption{New colouring $\phi^{'}$} \label{Figure:NewColouring}
\end{figure}
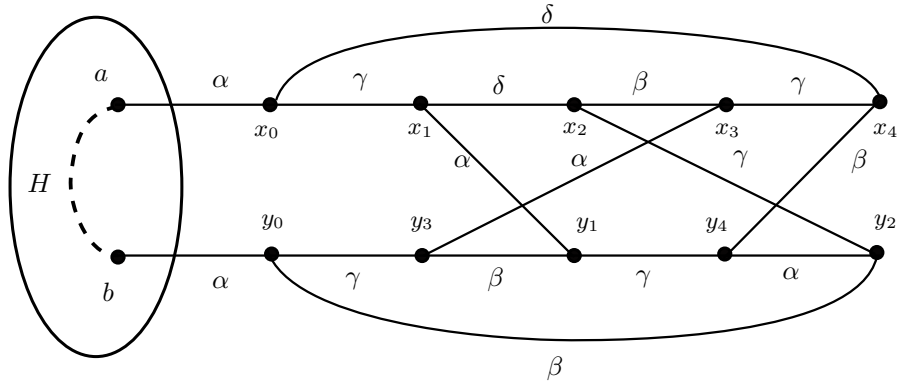

Let ${\mathcal C^{'}}$ be the set of odd cycles associated to edges coloured with $\delta$ by $\phi^{'}$.
Since we do not have change any colour in $H$, we have ${\mathcal C^{'}}=\mathcal C - \{C_{i},C_{j}\} +\{C^{'}_{i},C^{'}_{j}\}$
where $C^{'}_{i}$ is the chordless cycle of length $5$ $x_{1}x_{2}x_{3}y_{3}y_{1}$ and $C^{'}_{j}$ is the odd cycle obtained by
concatenation of the path $ax_{0}x_{4}y_{4}y_{2}y_{0}b$ and the path coloured alternately $\beta$ and $\alpha$ joining $a$ to $b$
in $H$ (dashed line in Figure \ref{Figure:NewColouring}), whose existence comes from Lemma \ref{Lemma:OddCycleAssociated}. Since
the length of $C^{'}_{j}$ is at least $7$, $l(\mathcal C^{'}) > l(\mathcal C)$, a contradiction with the initial choice of $\phi$.

\end{prf}

\ignore{\section{Application to bridgeless cubic graphs}

In this section we show that for any bridgeless cubic graph $G$
distinct from the Petersen graph we have $\gamma(G) \geq \frac{15
\strut}{\strut 17}$.

A {\em triangle} $T=\{a,b,c\}$ is said to be {\em reducible}
whenever its neighbours are distinct. When $T$ is a reducible
triangle in $G$ ($G$ having maximum degree $3$)  we can obtain a new
graph $G'$ with maximum degree $3$ by shrinking this triangle into a
single vertex and joining this new vertex to the neighbours of $T$ in
$G$.

\begin{lem} \label{Lemma:NoReducibleTriangle} \cite{AlbHaa96} Let $G$ be a graph with maximum degree $3$.
Assume that $T=\{a,b,c\}$ is a reducible triangle and let $G'$ be
the graph obtained by reduction of this triangle. Then
 $\gamma(G) > \gamma(G')$.
\end{lem}

Let $P_{12}$ be the cubic graph obtained from the Petersen graph by replacing a vertex by a triangle. One can easily verify
that  $s(P_{12})=2$  leading immediately to

\begin{lem} \label{Lemma:Petersen12}
 $\gamma(P_{12}) =\frac{8}{9}$.
\end{lem}

Let  $D_{k}$ ($k \geq 1$) be the graph depicted in Figure \ref{Figure:Diamondk} (where $a_{k}$ and $b_{k}$ are not adjacent)
and let $G$ be a  graph of maximum degree $3$  containing a subgraph $H$ isomorphic to $D_{k}$ for some $k \geq 1$. If we delete
the vertices of $H$ distinct from $a_{k}$ and $b_{k}$ and add a new edge between these two vertices we get a new  graph $G^{'}$
with maximum degree $3$. Let us say that $G^{'}$ is obtained from $G$ by {\em reducing} $D_{k}$

\begin{lem} \label{Lemma:NoDiamondk}  Let $G$ be a graph with maximum degree $3$.
Assume that $G$ contains a subgraph isomorphic to $D_{k}$ for some $k \geq 1$ and let $G'$ be
the graph obtained from $G$ by reducing $D_{k}$. Then $s(G)=s(G^{'})$ and
 $\gamma(G) > \gamma(G')$.
\end{lem}
\begin{prf}
Let $\phi$ be a \dm of $G$.Let $H$ be a subgraph of $G$ isomorphic to $D_k$. If $\phi(a_{k}a_{k-1})=\phi(b_{k}b_{k-1})$, then we get an immediate edge colouring
of $G^{'}$ by giving this common colour to the edge $a_{k}b_{k}$ which gives  $s(G^{'}) \leq s(G)$. If $\phi(a_{k}a_{k-1}) \neq \phi(b_{k}b_{k-1})$,
then one can easily verify that one edge at least of $D_{k}$ must be coloured with $\delta$. Hence we can obtain an  edge colouring of $G^{'}$ by
giving the colour $\delta$ to  $a_{k}b_{k}$ which gives  $s(G^{'}) \leq s(G)$ leading also to $s(G^{'}) \leq s(G)$.  Conversely, one can easily extend
a \dm of $G^{'}$ to a \dm of $G$ using at most $s(G^{'})$ edges coloured with $\delta$. Hence $s(G)=s(G^{'})$ as claimed.

Since $|E(G)| > |E(G^{'})|$, we have, by Lemma \ref{Lemma:bG}, $$\gamma(G)=1- \frac{s(G) \strut}{\strut |E(G)|} > 1- \frac{s(G^{'}) \strut}{\strut |E(G^{'})|}=\gamma(G^{'})$$.
\end{prf}

\begin{lem} \label{Lemma:Petersen_Diamondk}  Let $G$ be a bridgeless cubic graph and suppose that $G$ contains a subgraph isomorphic
to $D_{k}$ for some $k \geq 1$. Let $G'$ be
the graph obtained from $G$  by reducing $D_{k}$ and assume that $G^{'}$ is isomorphic to the Petersen graph. Then
 $\gamma(G) > \frac{15 \strut}{\strut 17}$.
\end{lem}

\begin{prf}
Obviously, $G$ contains at least $6$ edges more than the Petersen
graph. Since $\gamma(G)=1-\frac{s(G)}{|E(G)|}$ by Lemma
\ref{Lemma:bG} and $s(G)=2$ by Lemma \ref{Lemma:NoDiamondk}, we have
immediately $\gamma(G) \geq 1-\frac{2 \strut}{\strut 15+6} >
\frac{15 \strut}{\strut 17}$, as claimed.
\end{prf}

\begin{figure}[htb]
\begin{center}

\begin{center}
\scalebox{1} 
{
\begin{pspicture}(0,-1.638125)(11.102813,1.638125)
\definecolor{color2717b}{rgb}{0.03137254901960784,0.023529411764705882,0.023529411764705882}
\pscircle[linewidth=0.04,dimen=outer,fillstyle=solid,fillcolor=color2717b](0.4409375,1.0596875){0.1}
\pscircle[linewidth=0.04,dimen=outer,fillstyle=solid,fillcolor=color2717b](0.4409375,-0.9603125){0.1}
\pscircle[linewidth=0.04,dimen=outer,fillstyle=solid,fillcolor=color2717b](2.4409375,1.0596875){0.1}
\pscircle[linewidth=0.04,dimen=outer,fillstyle=solid,fillcolor=color2717b](2.4409375,-0.9403125){0.1}
\pscircle[linewidth=0.04,dimen=outer,fillstyle=solid,fillcolor=color2717b](6.4409375,1.0596875){0.1}
\pscircle[linewidth=0.04,dimen=outer,fillstyle=solid,fillcolor=color2717b](8.440937,1.0596875){0.1}
\pscircle[linewidth=0.04,dimen=outer,fillstyle=solid,fillcolor=color2717b](10.280937,0.1396875){0.1}
\pscircle[linewidth=0.04,dimen=outer,fillstyle=solid,fillcolor=color2717b](6.4409375,-0.9403125){0.1}
\pscircle[linewidth=0.04,dimen=outer,fillstyle=solid,fillcolor=color2717b](8.420938,-0.9203125){0.1}
\pscircle[linewidth=0.04,dimen=outer,fillstyle=solid,fillcolor=color2717b](9.220938,0.1396875){0.1}
\psline[linewidth=0.03cm](0.4209375,1.0396875)(8.440937,1.0396875)
\psline[linewidth=0.03cm](0.4409375,-0.9603125)(8.440937,-0.9603125)
\usefont{T1}{ptm}{m}{n}
\rput(0.40234375,1.4446875){$a_k$}
\usefont{T1}{ptm}{m}{n}
\rput(0.49234375,-1.4153125){$b_k$}
\psline[linewidth=0.03cm](2.4409375,1.0396875)(2.4409375,-0.9603125)
\psline[linewidth=0.03cm,linestyle=dotted,dotsep=0.16cm](3.0009375,0.0796875)(6.1609373,0.0796875)
\psline[linewidth=0.03cm](6.4409375,1.0796875)(6.4409375,-0.9603125)
\psline[linewidth=0.03cm](8.4609375,1.0596875)(9.240937,0.1396875)
\psline[linewidth=0.03cm](9.200937,0.1596875)(8.400937,-0.9603125)
\psline[linewidth=0.03cm](10.260938,0.1596875)(8.400937,-0.9203125)
\psline[linewidth=0.03cm](8.4609375,1.0196875)(10.340938,0.1196875)
\psline[linewidth=0.03cm](9.200937,0.1396875)(10.280937,0.1396875)
\usefont{T1}{ptm}{m}{n}
\rput(2.7723436,1.4446875){$a_{k-1}$}
\usefont{T1}{ptm}{m}{n}
\rput(2.6023438,-1.3553125){$b_{k-1}$}
\usefont{T1}{ptm}{m}{n}
\rput(6.382344,1.4446875){$a_1$}
\usefont{T1}{ptm}{m}{n}
\rput(8.342343,1.4446875){$a_0$}
\usefont{T1}{ptm}{m}{n}
\rput(8.452344,-1.3553125){$b_0$}
\usefont{T1}{ptm}{m}{n}
\rput(6.3923435,-1.3553125){$b_1$}
\usefont{T1}{ptm}{m}{n}
\rput(8.522344,0.2046875){$c$}
\usefont{T1}{ptm}{m}{n}
\rput(10.792344,0.1846875){$d$}
\end{pspicture} 
}
\end{center}
\end{center}
\caption{$D_k$} \label{Figure:Diamondk}
\end{figure}

\begin{thm} \label{Theorem:BrigelessCubic}Let $G$ be a bridgeless cubic graph distinct from the Petersen graph.
Then $\gamma(G) \geq \frac{15 \strut}{\strut 17}$
\end{thm}
\begin{prf}
When $G$ has at most $10$ vertices, it is well known that either $G$
is $3-$edge colourable  or $G$ is isomorphic to the Petersen graph.
The latter case is  is excluded by the hypothesis. When $G$ has $12$
vertices, the only bridgeless non $3-$edge colourable cubic graph is
$P_{12}$ for which the result is true by Lemma
\ref{Lemma:Petersen_Diamondk}. Hence, for bridgeless cubic graph
with at most $12$ vertices the result holds true. Assume by
induction that every bridgeless cubic graph $H$ with at most $ n\geq
12$ vertices is such that $\gamma(H) \geq \frac{15 \strut}{\strut
17}$ and let us prove the result for a bridgeless cubic graphs $G$
with $n+2$ vertices.

If $g_{odd}(G) \geq 5$ the result comes from Theorem \ref{Theorem:Principal}. We can thus suppose that $G$ contains a triangle
$T$. If this triangle is reducible, the result follows from Lemma \ref{Lemma:NoReducibleTriangle}. When $T$ is not reducible,
that means that either $G$ is reduced to $K_{4}$ (the three neighbours of $T$ are reduced to a single vertex) or $G$ contains
a subgraph isomorphic to $D_{k}$ for some $k\geq 1$. The former case is impossible since $G$ has at least $14$ vertices. In
the latter case, we use Lemma \ref{Lemma:NoDiamondk} or Lemma \ref{Lemma:Petersen_Diamondk} to conclude.
\end{prf}

}

\bibliographystyle{amsplain}
\bibliography{Parcimonious_7_8}
\end{document}